\documentclass[runningheads]{llncs}
\usepackage{graphicx}
\usepackage{xcolor}
\usepackage{multirow}
\usepackage{tikz}
\usepackage{algorithm,algpseudocode}
\usepackage{caption}
\usepackage{subcaption}
\usepackage{siunitx}
\usepackage{url}
\usepackage{makecell}
\usepackage{verbatim}

\def\checkmark{\tikz\fill[scale=0.26](0,.26) -- (.26,0) -- (1,.7) -- (.25,.15) -- cycle;} 

\begin{document}
\title{DIPPM: a Deep Learning Inference Performance Predictive Model using Graph Neural Networks}
\titlerunning{DIPPM: a Deep Learning Inference Performance Predictive Model}
%
\author{Karthick Panner Selvam  \and Mats Brorsson }

\authorrunning{K. Panner Selvam et al.}

\institute{SnT, University of Luxembourg, Luxembourg
\email{\{karthick.pannerselvam,mats.brorsson\}@uni.lu}}
\maketitle              
\begin{abstract}

Deep Learning (DL) has developed to become a corner-stone in many everyday applications that we are now relying on.  However, making sure that the DL model uses the underlying hardware efficiently takes a lot of effort. Knowledge about inference characteristics can help to find the right match so that enough resources are given to the model, but not too much.  
We have developed a DL Inference Performance Predictive Model (DIPPM) that predicts the inference \textit{latency}, \textit{energy}, and \textit{memory usage} of a given input DL  model on the NVIDIA A100 GPU. We also devised an algorithm to suggest the appropriate A100 Multi-Instance GPU profile from the output of DIPPM. 
We developed a methodology to convert DL models expressed in multiple frameworks to a generalized graph structure that is used in DIPPM. It means DIPPM can parse input DL models from various frameworks. Our DIPPM can be used not only helps to find suitable hardware configurations but also helps to perform rapid design-space exploration for the inference performance of a model.
We constructed a graph multi-regression dataset consisting of 10,508 different DL models to train and evaluate the performance of DIPPM, and reached a resulting Mean Absolute Percentage Error (MAPE) as low as 1.9\%.

\end{abstract}
\section{Introduction}
\label{intro}

Many important tasks a now relying on Deep learning models, for instance in computer vision and natural language processing domains~\cite{10.5555/3495724.3495883,9710580}. In recent years, researchers have focused on improving the efficiency of deep learning models to reduce the computation cost, energy consumption and increase the throughput of them without losing their accuracy. At the same time, hardware manufacturers like NVIDIA increase their computing power. For example, the NVIDIA A100\footnote{\url{https://www.nvidia.com/en-us/data-center/a100/}} GPU half-precision Tensor Core can perform matrix operations at 312 TFLOPS. But all deep learning models will not fully utilize the GPU because the workload and number of matrix operations will vary according to the problem domain. For this reason, NVIDIA created the Multi-Instance GPU (MIG\footnote{\url{https://docs.nvidia.com/datacenter/tesla/mig-user-guide/}}) technology starting from the Ampere architecture; they split the single physical GPU into multi-isolated GPU instances, so multiple applications can simultaneously run on different partitions of the same GPU, which then can be used more efficiently.

However, determining the DL model’s efficiency on a GPU is not straightforward. If we could predict parameters such as inference latency, energy consumption, and memory usage, we would not need to measure them on deployed models which is  a tedious and costly process. The predicted parameters could then also support efficient Neural Architecture Search (NAS)~\cite{10.5555/3322706.3361996}, efficient DL model design during development, and avoid job scheduling failures in data centers. According to Gao et al.~\cite{gao_estimating_2020}, most failed deep learning jobs in data centers are due to out-of-memory errors. 

\begin{figure}[t]
\centering
\includegraphics[width=0.8\textwidth]{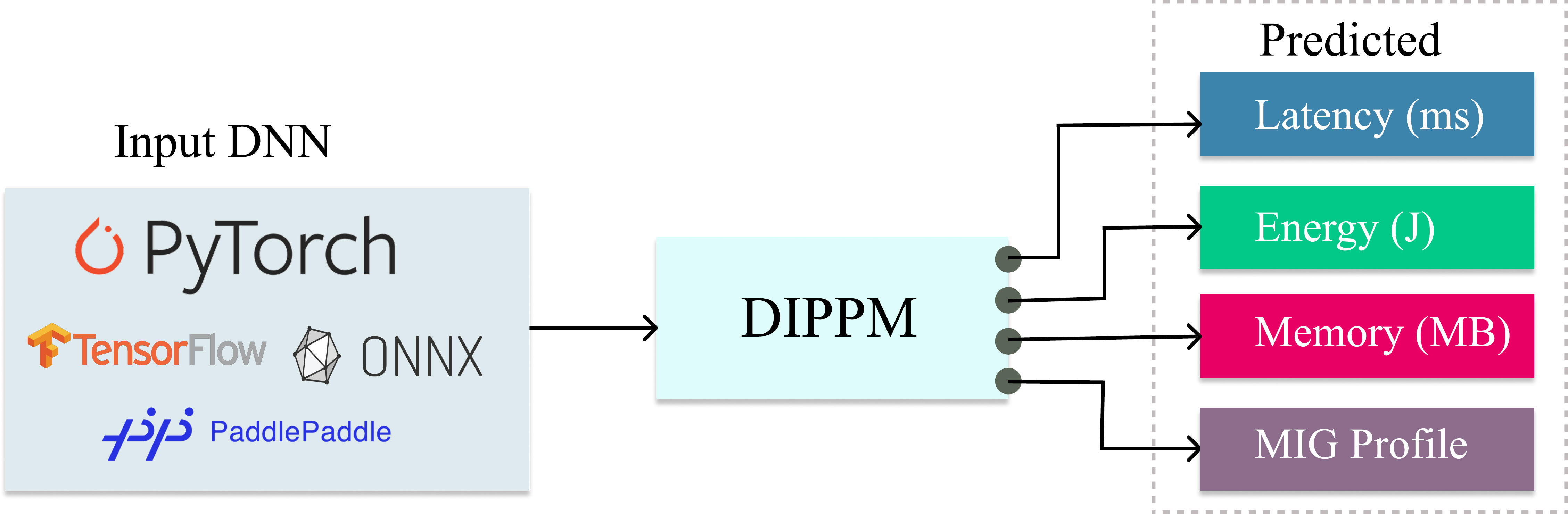}
\caption{DIPPM can predict the Latency, Energy, Memory requirement, and MIG Profile for inference on an NVIDIA A100 GPU without actually running on it. } 
\label{ppm_overview}
\end{figure}

In order to meet this need, we have developed a novel \textit{Deep Learning Inference Performance Predictive Model} (DIPPM) to support DL model developers in matching their models to the underlying hardware for inference. As shown in figure~\ref{ppm_overview}, DIPPM takes a deep learning model expressed in any of the frameworks: PyTorch, PaddlePaddle, Tensorflow, or ONNX, and will predict the latency (ms), energy (J), memory requirement (MB), and MIG profile for inference on an Nvidia A100 GPU without running on it. At the moment, the model is restricted to inference and the Nvidia A100 architecture, but we aim to relax these restrictions in future work. As far as we are aware, this is the first predictive model that can take input from any of the mentioned frameworks and to predict all of the metrics above.

Our contributions include the following:

\begin{itemize}

\item We have developed, trained and evaluated a performance predictive model which predicts inference latency, energy, memory, and MIG profile for A100 GPU with high accuracy.
\item We have developed a methodology to convert deep learning models from various deep learning frameworks into generalized graph structures for graph learning tasks in our performance predictive model. 
\item We have devised an algorithm to suggest the MIG profile from predicted Memory for the given input DL model. 
\item We have created an open-sourced performance predictive model dataset containing 10,508 graphs for graph-level multi-regression problems. 
\end{itemize}

Next, we discuss our work in relation to previous work in this area before presenting our methodology, experiments, and results.

\section{Related Work}
\label{related_work}

Performance prediction of deep learning models on modern architecture is a rather new research field being attended to only since a couple of years back. Bouhali et al.~\cite{10.1145/3444950.3447284} and Lu et al.~\cite{8863962} have carried out similar studies where a classical Multi-Layer Perceptron (MLP) is used to predict the inference latency of a given input DL model. Their approach was to collect high-level DL model features such as batch size, number of layers, and the total number of floating point operations (FLOPS) needed. They then fed these features into an MLP regressor as input to predict the latency of the given model. Bai et al.~\cite{DBLP:journals/corr/abs-2205-12095} used the same MLP method but predicted both the latency and memory. However, the classical MLP approach did not work very well due to the inability to capture a detailed view of the given input DL model. 

To solve the above problems, some researchers came up with a kernel additive method; they predict each kernel operation, such as convolution, dense, and LSTM, individually and sum up all kernel values to predict the overall performance of the DL model~\cite{8622396,DBLP:conf/iclr/QiST17,electronics11152316,10.1145/3477133.3477137,yang_perfestimator_2021,10.1145/3458864.3467882}. Yu et al.~\cite{yu_habitat_2021}  used the wave-scaling technique to predict the inference latency of the DL model on GPU, but this technique requires access to a GPU in order to make the prediction.

Kaufman et al. and Dudziak et al.~\cite{MLSYS2021_85d8ce59,10.5555/3495724.3496603} used graph learning instead of MLP to predict each kernel value. Still, they used the kernel additive method for inference latency prediction. However, this kernel additive method did not capture the overall network topology of the model, and instead it will affect the accuracy of the prediction. To solve the above problem, Liu et al.~\cite{10.1145/3545008.3545051} used a Graph level task to generalize the entire DL model into node embeddings and predicted the inference latency of the given DL model. However, they did not predict other parameters, such as memory usage and energy consumption.  

Li et al.~\cite{10.1145/3542929.3563510} tried to predict the MIG profiles on A100 GPU for the DL models. However, their methodology is not straightforward; they used CUDA Multi-Process Service (MPS) values to predict the MIG, So the model must run at least on the target hardware once to predict the MIG Profile. 

Most of the previous research work concentrated on parsing the input DL model from only one of the following frameworks (PyTorch, TensorFlow, PaddlePaddle, ONNX). As far as we are aware, none of the previous performance prediction models predicted Memory usage, Latency, Energy, and MIG profile simultaneously.

Our novel Deep Learning Inference Performance Predictive Model (DIPPM) fills a gap in previous work; a detailed comparison is shown in Table~\ref{tab:related_work}. DIPPM takes a deep learning model as input from various deep learning frameworks such as PyTorch, PaddlePaddle, TensorFlow, or ONNX and converts it to generalize graph with node features. We used a graph neural network and MIG predictor to predict the inference latency (ms), energy (J), memory (MB), and MIG profile for A100 GPU without actually running on it. 

\begin{table}[t]
\begin{minipage}{\textwidth}
\centering
\caption{Related Work comparison}
\label{tab:related_work}
\begin{tabular}{|l|c|c|c|c|c|c|c|}
\hline
\textbf{Related Works} & \textbf{A100} & \textbf{MIG} & \textbf{GNN\footnote{Using Graph Neural Network for performance prediction}} & \textbf{Multi-SF\footnote{Able to parse DL model expressed in Multiple DL Software Framework}} & \textbf{Latency} & \textbf{Power} & \textbf{Memory} \\ \hline
Ours (\textbf{DIPPM})           & \checkmark & \checkmark & \checkmark & \checkmark & \checkmark & \checkmark  & \checkmark \\ \hline
Bai et al.~\cite{DBLP:journals/corr/abs-2205-12095}     & - & - & - & - & \checkmark & -  & \checkmark \\ \hline
Bouhali et al.~\cite{10.1145/3444950.3447284} & - & - & - & - & \checkmark & -  & - \\ \hline
Dudziak et al.~\cite{10.5555/3495724.3496603} & - & - & \checkmark & - & \checkmark & -  & - \\ \hline
Justus et al.~\cite{8622396}  & - & - & - & - & \checkmark & -  & - \\ \hline
Kaufman et al.~\cite{MLSYS2021_85d8ce59}   & - & - & \checkmark & - & \checkmark & -  & - \\ \hline
Li et al.~\cite{10.1145/3542929.3563510}  & \checkmark & \checkmark & - & - & - & - & - \\ \hline
Liu et al.~\cite{10.1145/3545008.3545051}     & - & - & \checkmark & - & \checkmark & - & - \\ \hline
Lu et al.~\cite{8863962}      & - & - & - & - & \checkmark & \checkmark  & \checkmark \\ \hline
Qi et al.~\cite{DBLP:conf/iclr/QiST17} & - & - & - & - & \checkmark & -  & - \\ \hline
Sponner et al.~\cite{electronics11152316} & \checkmark & - & - & - & \checkmark & \checkmark  & \checkmark \\ \hline
Wang et al.~\cite{10.1145/3477133.3477137}    & - & - & - & - & \checkmark & -  & - \\ \hline
Yang et al.~\cite{yang_perfestimator_2021}    & - & - & - & - & \checkmark & -  & - \\ \hline
Yu et. al.~\cite{yu_habitat_2021}     & \checkmark & - & - & - & \checkmark & -  & - \\ \hline
Zhang et al.~\cite{10.1145/3458864.3467882}   & - & - & - & - & \checkmark & -  & - \\ \hline
\end{tabular}
\end{minipage}
\end{table}


\section{Methodology}

\begin{figure}[t]
\centering
\includegraphics[width=\textwidth]{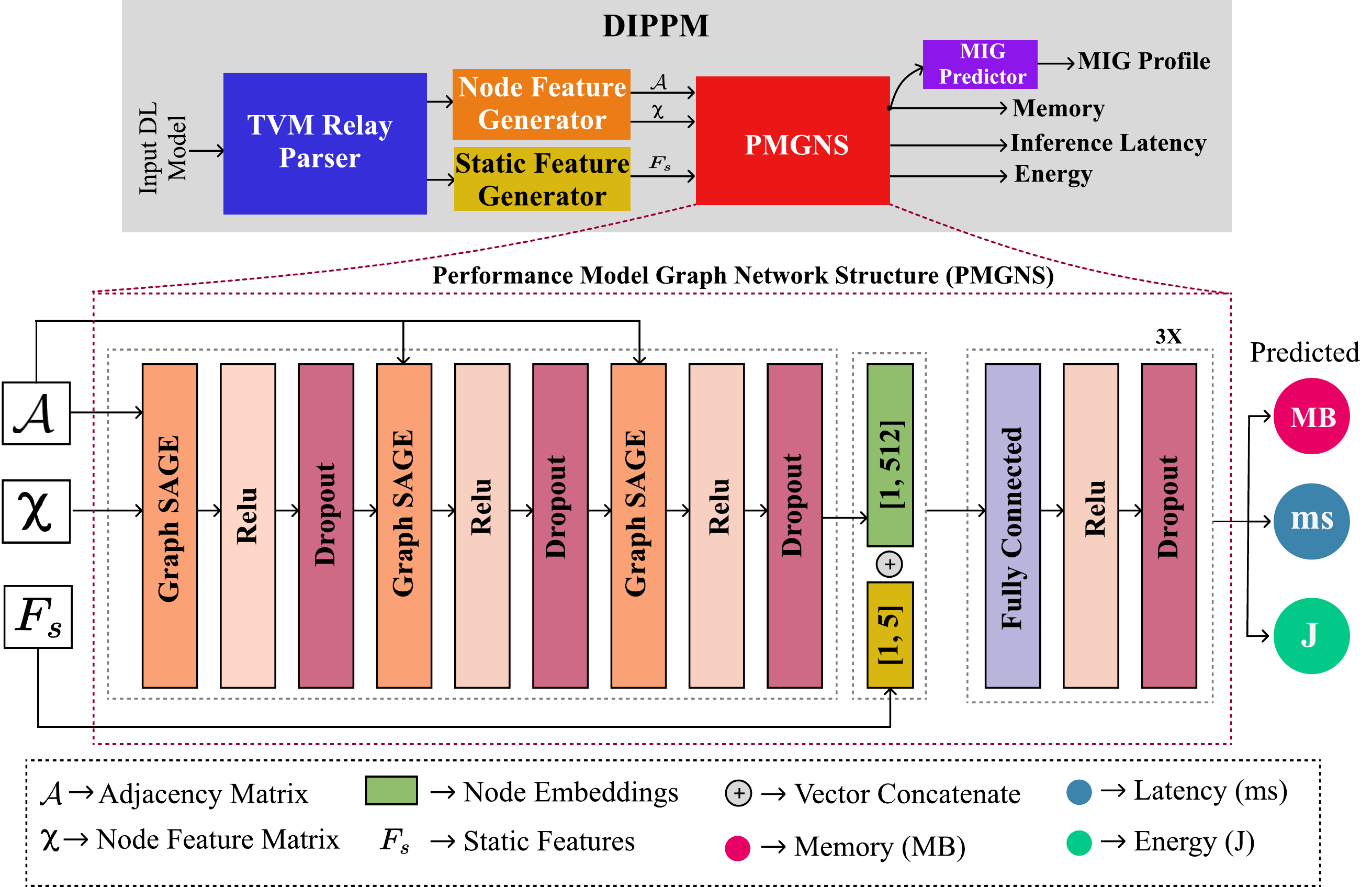}
\caption{Overview of DIPPM Architecture} 
\label{dippm}
\end{figure}

The architecture of DIPPM consists of five main components: Relay Parser, Node Feature Generator, Static Feature Generator, Performance Model Graph Network Structure (PMGNS), and MIG Predictor, as shown in Fig.~\ref{dippm}. We will explain each component individually in this section.

\subsection{Relay Parser}
\label{sec:relay-parser}

The Relay Parser takes as input a DL model expressed in one of several supported DL frameworks, converts it to an Intermediate Representation (IR), and passes this IR into the Node Feature Generator and the Static Feature Generator components.

Most of the previously proposed performance models are able to parse the given input DL model from a single DL framework, not from several, as we already discussed in Section~\ref{related_work}. To enable the use of multiple frameworks, we used a relay, which is a high-level IR for DL models~\cite{10.1145/3211346.3211348}. It has been used to compile DL models for inference in the TVM\footnote{\url{https://tvm.apache.org/}} framework. We are inspired by the way they convert the DL model from various DL frameworks into a high-level IR format and therefore used their technique in our DIPPM architecture. It allows parsing given input DL models from various frameworks, but we have chosen to limit ourselves to PyTorch, TensorFlow, ONNX, and PaddlePaddle. We pass this DL IR to the subsequent components in our DIPPM architecture.  

\subsection{Node Feature Generator}
\label{node_feature_generator}

The Node Feature Generator (NFG) converts the DL IR into  an Adjacency Matrix ($\mathcal A$) and a Node feature matrix ($\mathcal X$) and passes this data to the PMGNS component.

The NFG takes the IR from the relay parser component. The IR is itself a computational data flow graph containing more information than needed for our performance prediction. Therefore we filter and pre-process the graph by post-order graph traversal to collect necessary node information. The nodes in the IR contain useful features such as operator name, attributes, and output shape of the operator, which after this first filtering step are converted into a suitable data format for our performance prediction. In the subsequent step, we loop through the nodes and, for each operator node, generate node features with a fixed length of 32. 

The central part of the NFG is to generate an \textbf{Adjacency Matrix} ($\mathcal A$) and a \textbf{Node feature matrix} ($\mathcal X$) as expressed in algorithm \ref{create_graph}. $\mathcal X$ has the shape of $[N_{op}, N_{features}]$, where $N_{op}$ is the number of operator nodes in the IR and $N_{features}$ is the number of features. 
In order to create node features $\mathcal F_{n}$ for each $node$, first, we need to encode the node operator name into a one hot encoding as can be seen on line \ref{onehot} in algorithm~\ref{create_graph}. Then extract the node attributes $\mathcal F_{attr}$ and output shape $\mathcal F_{shape}$ into vectors. Finally, perform vector concatenation to generate $\mathcal F_{n}$ for a node. We repeat this operation for each node and create the $\mathcal G$. From the  $\mathcal G$, we extract  $\mathcal A$,  $\mathcal X$ that are passed to the main part of our model, the Performance Model Graph Network Structure. 

\begin{algorithm} [t]
\caption{Algorithm to convert DL model IR into a graph with node features}
CreateGraph takes input IR and filters it by post-order traversal. Collect node features for each node and generate a new graph $\mathcal G$ with node features, finally extract node feature matrix $\mathcal X$ and adjacency matrix $\mathcal A$ from $\mathcal G$.
\label{create_graph}
\begin{algorithmic}[1]
\algnewcommand\algorithmicforeach{\textbf{for each}}
\algdef{S}[FOR]{ForEach}[1]{\algorithmicforeach\ #1\ \algorithmicdo}
\Function{CreatGraph}{$IR$}       \Comment{IR from Relay Parser Component}
    \State $\mathcal N \gets filter\_and\_preprocess(IR)$
    \State $\mathcal G \gets \emptyset$ \Comment{Create empty directed graph}
    \ForEach {$node \in \mathcal N $}  \Comment{where $node$ is node in node\_list  $\mathcal N$}
        \If{$node.op \in $ [operators]} \Comment{Check node is a operator}
        \State $\mathcal F_{oh} \gets one\_hot\_encoder(node.op)$  \label{onehot}
        \State $\mathcal F_{attr} \gets ExtractAttributes(node)$
        \State $\mathcal F_{shape} \gets ExtractOutshape(node)$
        \State $\mathcal F_{node}  \gets \mathcal F_{oh} \oplus \mathcal F_{attr}  \oplus \mathcal F_{shape}$ \label{node_feature}
        \State $\mathcal G.add\_node(node.parent, node.id, \mathcal F_{node})$ \Comment{Nodes are added in sequence}
        \EndIf
    \EndFor
     \State $\mathcal A \gets GetAdjacencyMatrix(\mathcal G)$
     \State $\mathcal X \gets GetNodeFeatureMatrix(\mathcal G)$
    \State \Return  $\mathcal A, \mathcal X$
\EndFunction
\end{algorithmic}
\end{algorithm}

\subsection{Static Feature Generator}
\label{static_generator}

The Static Feature Generator (SFG) takes the IR from the relay parser component and generates static features $\mathcal F_{s}$ for a given DL model and passes them into the graph network structure. 

For this experiment, we limited ourselves to five static features. First, we calculate the $\mathcal F_{mac}$ total multiply-accumulate (MACs) of the given DL model. We used the  TVM relay analysis API to calculate total MACs, but it is limited to calculating MACs for the following operators (in TVM notation): Conv2D, Conv2D transpose, dense, and batch matmul. Then we calculate the total number of convolutions $F_{Tconv}$, Dense $F_{Tdense}$, and Relu $F_{Trelu}$ operators from the IR. We included batch size $F_{batch}$ as one of the static features because it gives the ability to predict values for various batch sizes of a given model. Finally, we concatenate all the features into a vector $\mathcal F_{s}$ as expressed in equation~\ref{static_feature}. The feature set $\mathcal F_{s}$ is subsequently passed to the following graph network structure.

\begin{equation}
\label{static_feature}
    \mathcal F_{s} \gets \mathcal F_{mac}  \oplus \mathcal F_{batch} \oplus \mathcal F_{Tconv} \oplus \mathcal F_{Tdense} \oplus \mathcal F_{Trelu}
\end{equation}

\subsection{Performance Model Graph Network Structure (PMGNS)}
\label{pmgns}

The PMGNS takes the node feature matrix ($\mathcal X$), the adjacency matrix ($\mathcal A$) from the Node Feature Generator component, and the feature set ($\mathcal F_{s}$) from the Static feature generator and predicts the given input DL model's memory, latency, and energy, as shown in Fig.~\ref{dippm}.

The PMGNS must be trained  before prediction, as explained in section~\ref{experiment}. The core idea of the PMGNS is to generate the node embedding $z$ from $\mathcal X$ and $\mathcal A$ and then to perform vector concatenation of $z$ with $\mathcal F_{s}$. Finally, we pass the concatenated vector into a Fully Connected layer for prediction, as shown in Fig. \ref{dippm}. In order to generate $z$, we used the graphSAGE algorithm suggested by Hamilton et al.~\cite{10.5555/3294771.3294869}, because of its inductive node embedding, which means it can generate embedding for unseen nodes without pretraining. 

We already discussed that we generate node features of each node in the section~\ref{node_feature_generator}. The graphSAGE algorithm will convert node features into a node embedding $z$ which is more amenable for model training. 
The PMGNS contains three sequential graphSAGE blocks and three sequential Fully connected (FC) blocks as shown in Fig. \ref{dippm}. At the end of the final graphSAGE block, we get the generalized node embedding of given $\mathcal X$ and $\mathcal A$, which we concatenate with $\mathcal F_{s}$. Then we pass the concatenated vector into FC to predict the memory (MB), latency (ms), and energy (J).

\subsection{MIG Predictor}
The MIG predictor takes the memory prediction from PMGNS and predicts the appropriate MIG profile for a given DL model, as shown in Fig.~\ref{dippm}.

As mentioned in the introduction, the Multi-instance GPU (MIG) technology allows to split an A100 GPU into multiple instances so that multiple applications can use the GPU simultaneously. The different instances differ in their compute capability and, most importantly, in the maximum memory limit that is allowed to be used. The four MIG profiles of the A100 GPU that we consider here are: 1g.5gb, 2g.10gb, 3g.20gb, and 7g.40gb, where the number in front of "gb" denotes the maximum amount of memory in GB that the application can use on that instance. For example, the maximum memory limit of 1g.5gb is 5GB, and 7g.40gb is 40GB. 

\begin{figure}[t]
\centering
\includegraphics[width=\textwidth]{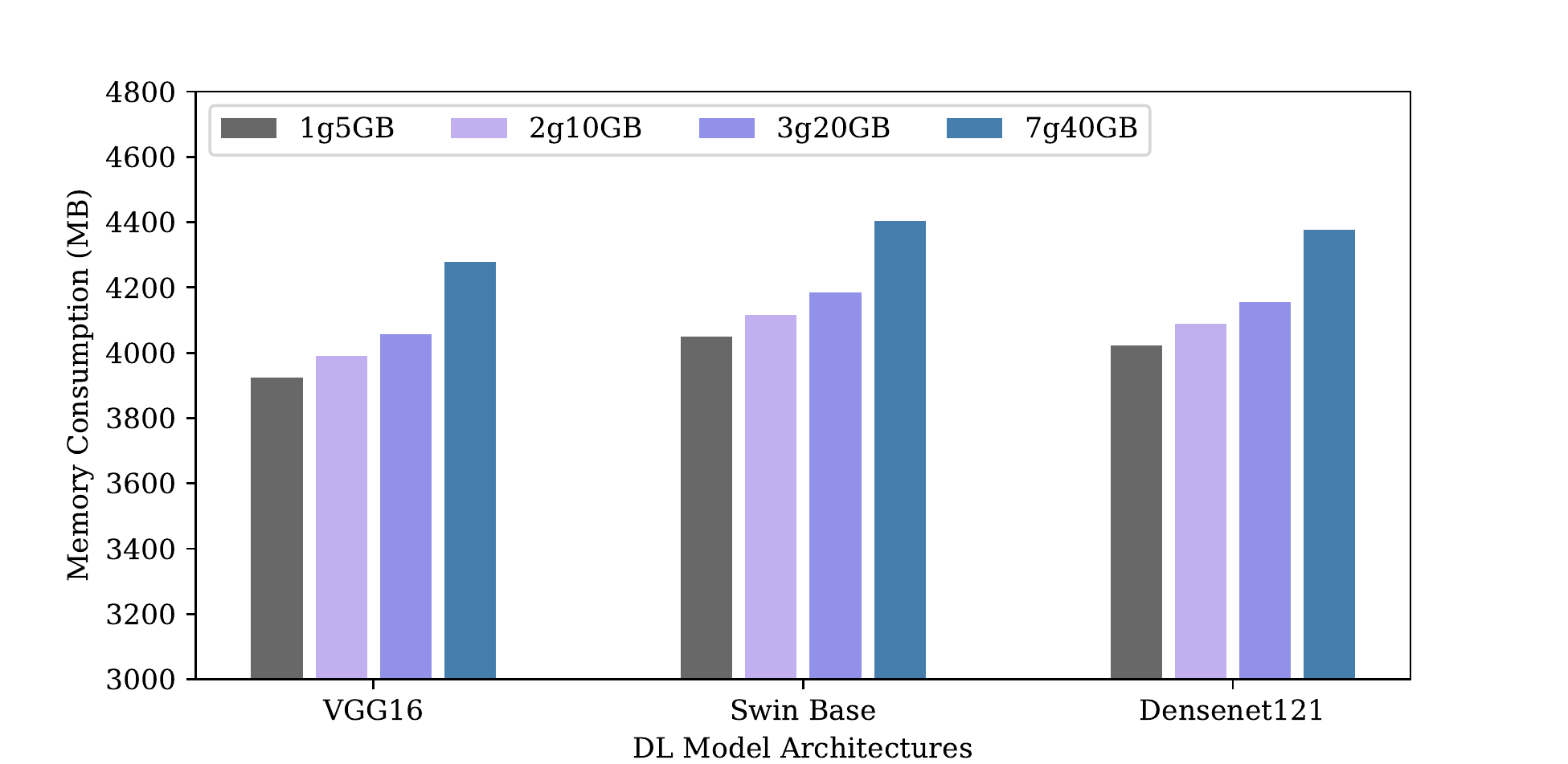}
\caption{MIG Profile comparison of three different DL models memory consumption on A100 GPU. We used batch size 16 for VGG16 and Densenet121 model and batch size 8 for Swin base model.} \label{mig}
\end{figure}

For a given input DL model, PMGNS predicts memory for 7g.40gb MIG profile, which is the full GPU.
We found that this prediction can be used as a pessimistic value to guide the choice of MIG profile. Fig.~\ref{mig} shows manual memory consumption measurements of the same DL model inference on different profiles. The results show no significant difference in the memory allocation of DL in the different MIG profiles even though the consumption slightly increases with the capacity of the MIG profile. The memory consumption is always the highest when running on the 7g.40gb MIG profile.

As mentioned, PMGNS predicts memory for 7g.40gb, so we claim that predicted memory will be an upper bound. Then we perform a rule-based prediction to predict the MIG profile for the given input DL model, as shown in equation \ref{migp}. Where $\alpha$ is predicted memory from PMGNS.

\begin{equation}
\label{migp}
\mathrm{MIG}({\alpha}) = \left\{ \begin{array}{ll} \textrm{1g.5gb}, & \mathrm{if} \ {0gb < {\alpha} < \textrm{5gb}} \\ \textrm{2g.10gb}, & \mathrm{if} \ {\textrm{5gb} <{\alpha} < \textrm{10gb}} \\ \textrm{3g.20gb}, & \mathrm{if} \ {\textrm{10gb} < {\alpha} < \textrm{20gb}} \\ \textrm{7g.40gb}, & \mathrm{if} \ {{\textrm{20gb} < \alpha} < \textrm{40gb}} \\ \mathrm{None}, & \mathrm{otherwise} \end{array} \right.
\end{equation}

\section{Experiments \& Results}
\label{experiment}

\subsection{The DIPPM Dataset}
\label{dataset}

We constructed a graph-level multi-regression dataset containing 10,508 DL models from different model families to train and evaluate our DIPPM. The dataset distribution is shown in Table~\ref{tab:dataset}. Why do we need to create our own dataset? To the best of our knowledge, the previous predictive performance model dataset doesn't capture  memory consumption, inference latency, and energy consumption parameters for wide-range DL models on A100 GPU.

Our dataset consists of DL models represented in graph structure, as generated by the Relay parser described in section~\ref{sec:relay-parser}. Each data point consists of  four variables: $\mathcal X$, $\mathcal A$, $\mathcal Y$, and $\mathcal F_{s}$, where $\mathcal X$ and $\mathcal A$ are the Node feature matrix and Adjacency Matrix, respectively, as discussed in section~\ref{node_feature_generator}, and $\mathcal F_{s}$ is the static features of the DL model as discussed in section~\ref{static_generator}. We used the Nvidia Management Library\footnote{\url{https://developer.nvidia.com/nvidia-management-library-nvml}} and the  CUDA toolkit\footnote{\url{https://developer.nvidia.com/cuda-toolkit}} to measure the energy, memory, and inference latency of each given model in the dataset. For each model, we ran the inference five times to warm up the architecture and then the inference 30 times, and then took the arithmetic mean of those 30 values to derive the $\mathcal Y$, where  $\mathcal Y$ consists of inference latency (ms), memory usage (MB), and energy (J) for a given DL on A100 GPU. 

We used a full A100 40GB GPU, or it is equivalent to using 7g.40gb MIG profile to collect all the metrics.

\renewcommand{\arraystretch}{1}
\setlength{\tabcolsep}{2pt}
\begin{table}[t]
\centering
\caption{DIPPM  Graph dataset distribution}
\label{tab:dataset}
\begin{tabular}{l|S|S}
\hline 
\textbf{Model Family} & \textbf{\# of Graphs} & \textbf{Percentage (\%)} \\ \hline \hline
Efficientnet & \num{1729} & 16.45 \\ \hline
Mnasnet      & \num{1001} & 9.53  \\ \hline
Mobilenet    & \num{1591} & 15.14 \\ \hline
Resnet       & \num{1152} & 10.96 \\ \hline
Vgg          & \num{1536} & 14.62 \\ \hline
Swin         & \num{547}  & 5.21  \\ \hline
Vit          & \num{520}  & 4.95  \\ \hline
Densenet     & \num{768}  & 7.31  \\ \hline
Visformer    & \num{768}  & 7.31  \\ \hline
Poolformer   & \num{896}  & 8.53  \\ \hline
\textbf{Total} & \textbf{\num{10508}}        & \num{100}\%   \\ \hline
\end{tabular}
\end{table}

\subsection{Enviroment setup}

We used an HPC cluster at the Jülich research centre in Germany called JUWELS Booster for our experiments\footnote{\url{https://apps.fz-juelich.de/jsc/hps/juwels/booster-overview.html}}. It is equipped with 936 nodes, each with AMD EPYC 7402 processors, 2 sockets per node, 24 cores per socket, 512 GB DDR4-3200 RAM and 4 NVIDIA A100 Tensor Core GPUs with 40 GB HBM. 

The main software packages used in the experiments are: Python 3.10, CUDA 11.7 torch 1.13.1, torch-geometric 2.2.0, torch-scatter 2.1.0, and torch-sparse 0.6.16.

\subsection{Evaluation}

The Performance Model Graph Network Structure is the main component in DIPPM, and we used the PyTorch geometric library to create our model, as shown in Fig.~\ref{dippm}. We split our constructed  dataset into three parts randomly: training set 70\%, validation set 15\%, and a test set 15\%. 

In order to validate that graphSAGE performs better than other GNN algorithms and plain MLP, we compared graphSAGE with the following other algorithms:, GAT~\cite{velickovic2018graph}, GCN~\cite{kipf2017semi}, GIN~\cite{xu2018how}, and finally, plain MLP without GNN. Table~\ref{tab:gnn-settings} summarizes the settings used. The learning rate was determined using a learning rate finder as suggested by Smith~\cite{7926641}. The Huber loss function achieved a higher accuracy than mean square error, which is why we chose that one.

\begin{table}[t]
\caption{Settings in GNN comparison.}
\label{tab:gnn-settings}
\centering
\begin{tabular}{l|l}
\hline
\textbf{Setting}      & \textbf{Value}  \\ \hline
\hline
Dataset partition     & Train (70\%) / Validation (15\%) / Test (15\%) \\ \hline
Nr hidden layers      & 512  \\ \hline
Dropout probability \hspace{3mm}  & 0.05  \\ \hline
Optimizer             & Adam  \\ \hline
Learning rate         & $ 2.754 \cdot 10^{-5}$ \\ \hline
Loss function         & Huber \\ \hline
\end{tabular}
\end{table}

\begin{table}[t]
\caption{Comparison with different GNN algorithms and MLP with graphSAGE, we trained all the models for 10 epochs and used Mean Average Percentage Error for validation. The results indicate that DIPPM with graphSAGE performs significantly better than other variants.}
\label{tab:ablation}
\centering
\begin{tabular}{l|c c c}
\hline
\textbf{Model}        & \textbf{Training} & \textbf{Validation} &\textbf{Test} \\ \hline \hline
GAT & 0.497 & 0.379 & 0.367 \\ \hline
GCN & 0.212 & 0.178 & 0.175 \\ \hline
GIN & 0.488 & 0.394 & 0.382 \\ \hline
MLP & 0.371 & 0.387 & 0.366 \\ \hline
\textbf{(Ours) GraphSAGE} & \textbf{0.182}  &  \textbf{0.159}     & \textbf{0.160}              \\ \hline
\end{tabular}
\end{table}

For the initial experiment, we trained for 10 epochs and used Mean Average Percentage Error (MAPE) as accuracy metric to validate DIPPM. A MAPE value close to zero indicates good performance on regression prediction. Table~\ref{tab:ablation} shows that graphSAGE gives a lower MAPE value in all of training, validation, and test datasets. Without using a GNN, MLP gives 0.366 of MAPE.  With graphSAGE, MAPE is 0.160 on the test dataset which is a significant improvement on a multi-regression problem. We conclude that graphSAGE outperforms other GNN algorithms, and MLP because of its inductive learning, as discussed in section~\ref{pmgns}. 

\begin{figure}[t]
     \centering
     \begin{subfigure}[l]{0.49\textwidth}
        \label{predictions:lat}
         \centering
        \includegraphics[width=\textwidth]{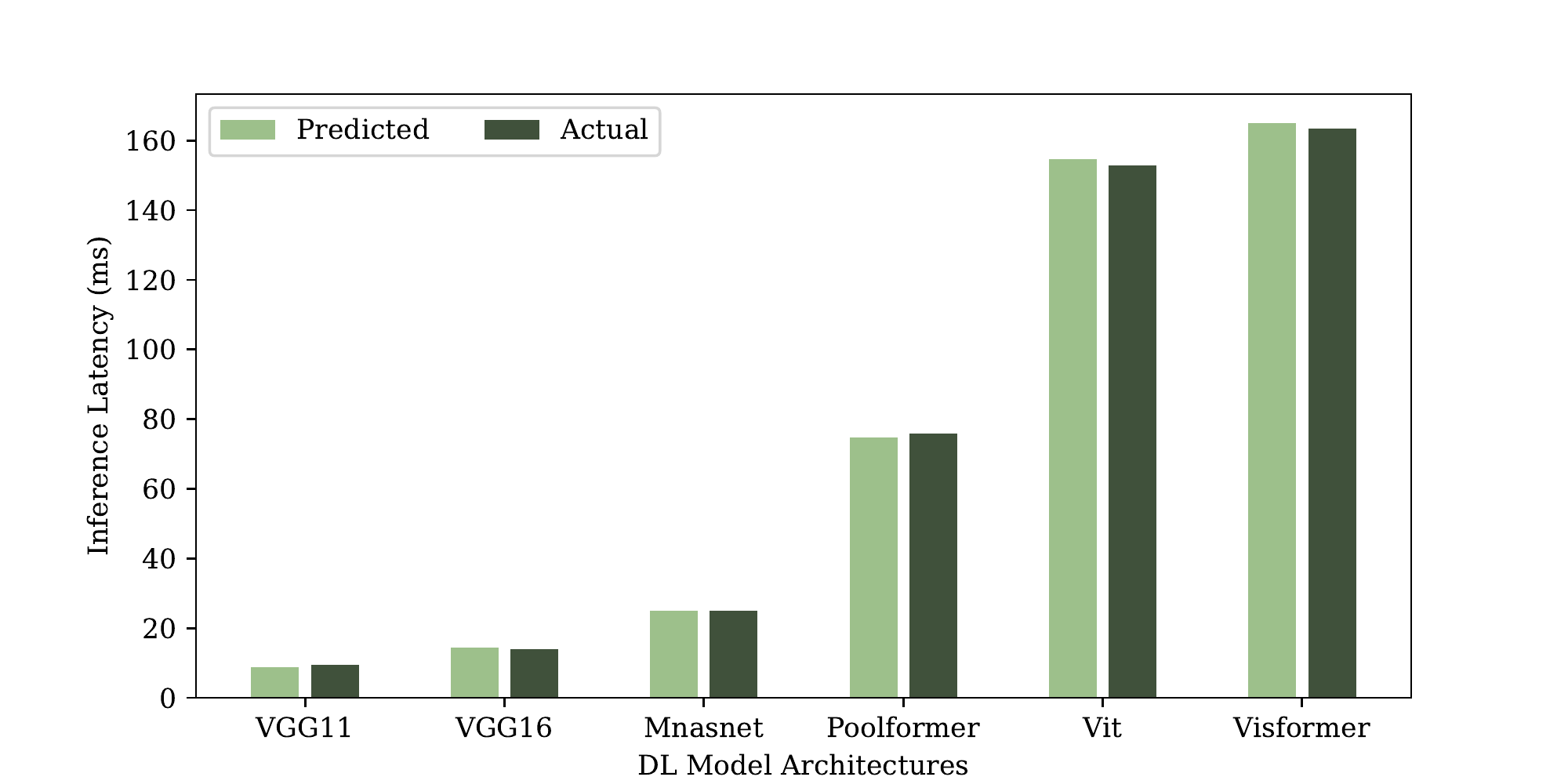}         \caption{Inference latency (ms).}
     \end{subfigure}
     \hfill
     \begin{subfigure}[r]{0.49\textwidth}
         \textbf{}
         \centering
        \includegraphics[width=\textwidth]{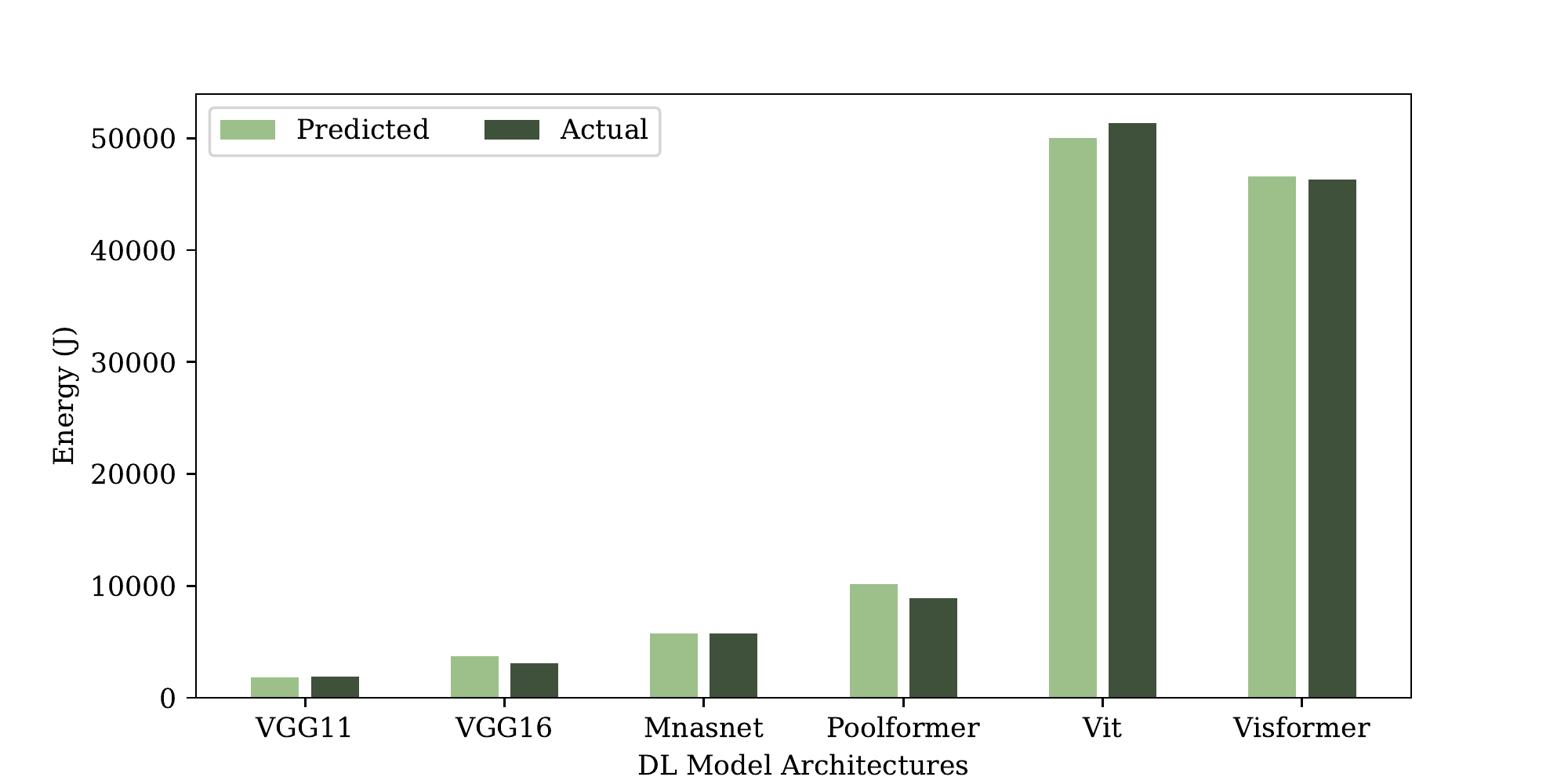}         \caption{Energy (J).}
     \end{subfigure}
          \hfill
     \begin{subfigure}[b]{0.49\textwidth}
         \centering
        \includegraphics[width=\textwidth]{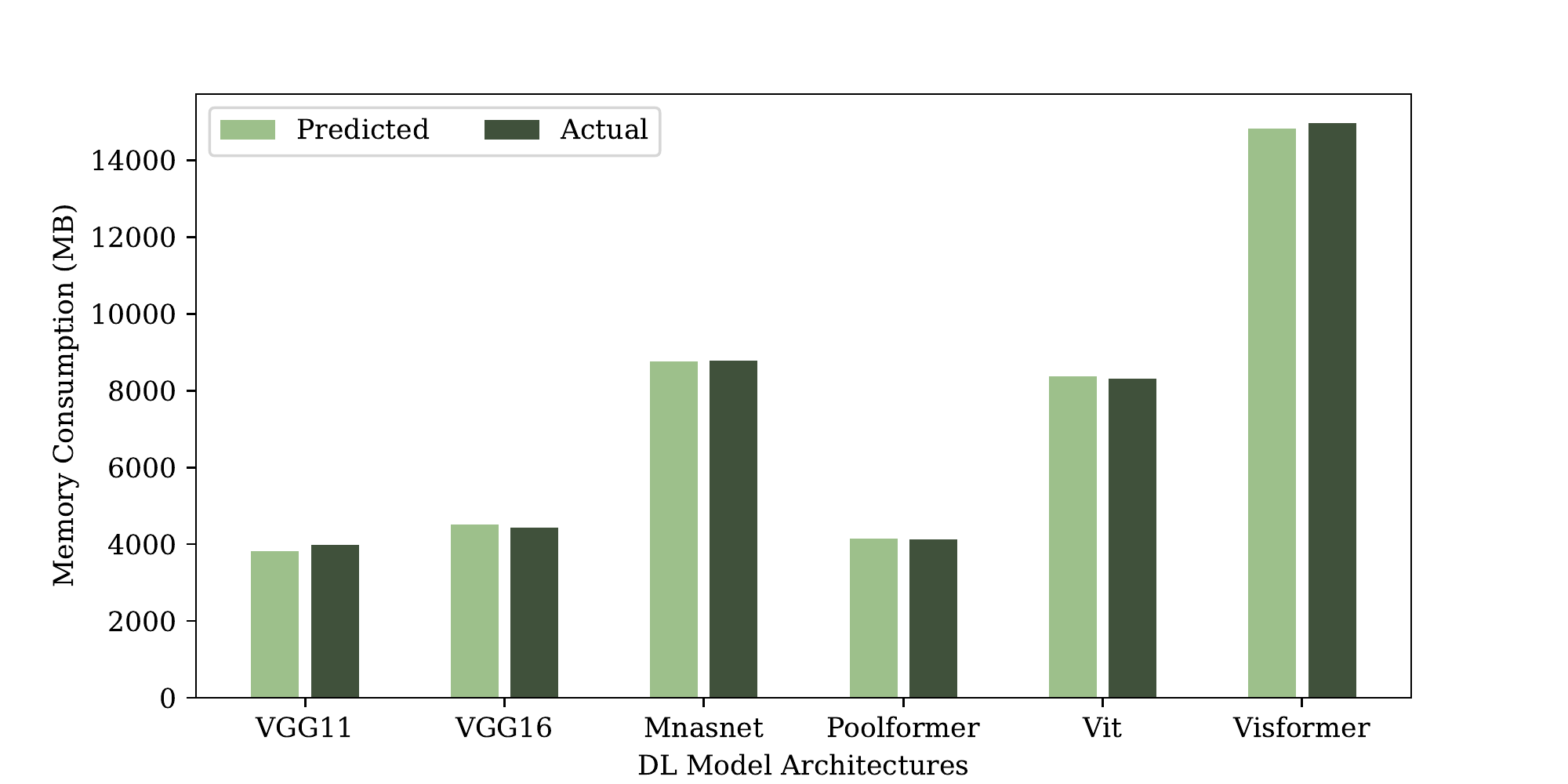}         \caption{Memory consumption (MB).}
     \end{subfigure}
\caption{Comparison of actual value with DIPPM predicted values on the test dataset. Results show that DIPPM predictions are close to the actual predictions.}
\label{predictions}
\end{figure}

After this encouraging result we increased the number of epochs for training our DIPPM with graphSAGE to increase the prediction accuracy. After 500 epochs, we attained MAPE of 0.041 on training and 0.023 on the validation dataset. In the end, we attained 1.9\% MAPE on the test dataset. Some of the DIPPM predictions on the test dataset are shown in Fig. \ref{predictions}.

\subsection{Prediction of MIG Profiles}

In order to verify the MIG profile prediction for a given DL model, we compared the actual MIG profile value with the predicted MIG profile from the DIPPM, as shown in table~\ref{tab:mig}. To calculate the actual suitable MIG profile, we divide actual memory consumption by the maximum memory limit of the MIG profiles. The higher the value is, the more appropriate profile for the given DL model. 

For example, the predicted memory consumption for densenet121 at batch size 8 is 2865 MB. The actual memory consumption for the 7g.40gb MIG profile is 3272 MB. You can see that our DIPPM correctly predicted the MIG profile 1g.5gb for densenet121. 

It is interesting to note that the densent121 models are from our test dataset and the swin base patch4 model is not in our DIPPM dataset but a similar swin base model family was used to train DIPPM. The convnext models are completely unseen to our DIPPM, but it's still predicting the MIG profile correctly. 

\begin{table}[t]
\centering
\caption{DIPPM MIG profile prediction for seen and unseen DL model architectures. (densenet*: seen, swin*: partially seen, convnext*: unseen). }
\label{tab:mig}
\begin{tabular}{l|c|cc|ccccc}
\hline
\multicolumn{1}{c|}{} &
   &
  \multicolumn{2}{c|}{\textbf{Predicted}} &
  \multicolumn{5}{c}{\textbf{Actual}} \\ \cline{3-9} 
\multicolumn{1}{c|}{\multirow{-2}{*}{\textbf{Model}}} &
  \multirow{-2.2}{*}{\thead{\textbf{Batch} \\ \textbf{size}}} &
  \textbf{MIG} &
  \textbf{Mem} &
  \textbf{Mem} &
  \textbf{1g.5gb} &
  \textbf{2g.10gb} &
  \textbf{3g.20gb} &
  \textbf{7g.40gb} \\ \hline \hline
densenet121 &
  8 & 1g.5gb & 2865 & 3272 & \textbf{58\%} & 30\% & 
15\% & 8\% \\ \hline
densenet121 &   32 &  2g.10gb &  5952 &6294 &  & \textbf{60\%} & 30\% & 16\%  \\ \hline
swin\_base\_patch4 &   2 & 1g.5gb &   2873 & 2944 & \textbf{52\%} & 27\% & 14\% & 7\%  \\ \hline
swin\_base\_patch4 &   16 & 2g.10gb &   6736 & 6156  &  &\textbf{59\% } & 30\%  & 15\% \\ \hline
convnext\_base &   4 &1g.5gb &  4771 & 1652 & \textbf{61\% } & 31\% & 16\%  & 8\% \\ \hline
convnext\_base &   128 & 7g.40gb &   26439 & 30996 & & & &   \textbf{77\%} \\ \hline
\end{tabular}
\end{table}

\subsection{DIPPM Usability aspects}

\begin{figure}[t]
\centering
\includegraphics[width=0.9\textwidth]{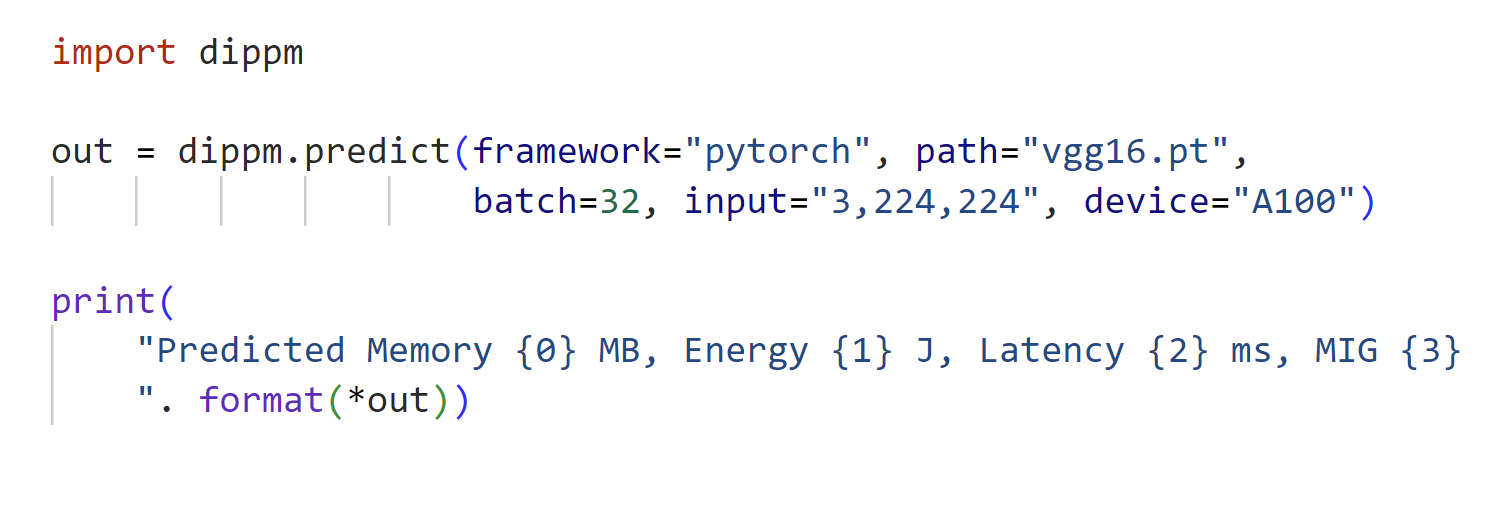}
\caption{A sample code to use DIPPM for performance prediction of VGG16 DL model developed by PyTorch framework.} 
\label{code}
\end{figure}

DIPPM takes basic parameters like frameworks, model path, batch, and input size, and finally, device type. As of now, we only considered A100 GPU; we are working to extend DIPPM to various hardware platforms. With a simple python API call, DIPPM predicts memory, latency, energy, and MIG profile for the given model, as can be seen in Fig.~\ref{code}.

\section{Conclusion}

We have developed a novel Deep Learning (DL) Inference Performance Predictive Model (DIPPM) to predict the inference latency, energy, and memory consumption of a given input DL model on an A100 GPU without running on it. Furthermore, We devised an algorithm to select the appropriate MIG profile from the memory consumption predicted by DIPPM. 

The model includes a methodology to convert the DL model represented in various frameworks to a generalized graph structure for performance prediction. To the best of our knowledge, DIPPM can help to develop an efficient DL model to utilize the underlying GPU effectively. Furthermore, we constructed and open-sourced\footnote{The URL to the dataset will be provided with the camera-ready version of the paper.} a multi-regression graph dataset containing 10,508 DL models for performance prediction. It can even be used to evaluate other graph-based multi-regression GNN algorithms. Finally, we achieved 1.89\% MAPE on our dataset.

\section*{Acknowledgment}
This work has been funded by EuroHPC JU under contract number 955513 and by the Luxembourg National Research Fund (FNR) under contract number 15092355.

\end{document}